\documentclass[preprint,12pt]{elsarticle}

\usepackage{amsmath}
\usepackage{amssymb}
\journal{Physica A}

\begin{document}

\begin{frontmatter}

\title{Mobility driven coexistence of living organisms}

\author[MAR]{B.F. de Oliveira}
\author[MAR]{M.V. de Moraes}
\author[PAR]{D. Bazeia}
\author[EN]{and A. Szolnoki}

\address[MAR]{Departamento de F\'\i sica, Universidade Estadual de Maring\'a, 87020-900 Maring\'a, PR, Brazil}
\address[PAR]{Departamento de F\'\i sica, Universidade Federal da Para\'\i ba, 58051-970 Jo\~ao Pessoa, PB, Brazil}
\address[EN]{Institute of Technical Physics and Materials Science, Centre for Energy Research, P.O.~Box 49, H-1525 Budapest, Hungary}

\begin{abstract}
We propose a minimal off-lattice model of living organisms where just a very few dynamical rules of growth are assumed. The stable coexistence of many clusters is detected when we replace the global restriction rule by a locally applied one. A rich variety of evolving patterns is revealed where players movement has a decisive role on the evolutionary outcome. For example, intensive individual mobility may jeopardize the survival of the population, but if we increase players movement further then it can save the population. Notably, the collective drive of population members is capable to compensate the negative consequence of intensive movement and keeps the system alive. When the drive becomes biased then the resulting unidirectional flow alters the stable pattern and produce a stripe-like state instead of the previously observed hexagonal arrangement of clusters. Interestingly, the rotation of stripes can be flipped if the individual movement exceeds a threshold value.
\end{abstract}
\begin{keyword}
\texttt clustering\sep\ drive\sep\ pattern formation
\end{keyword}

\end{frontmatter}

\section{Introduction}
\label{intro}

Clusters are essential and abundant patterns in wide variety of living systems ranging from the simplest organizations, such as bacterias, to more complex ones, like insects, fishes, or birds \cite{lui_nchem13,agarwala_fe93,camazine_01}. But mammals and even humans can also be observed as they form the mentioned shape \cite{vicsek_n10}. Evidently, there are several and often independent reasons why clusters emerge and maintain because external conditions and governing dynamics may differ significantly \cite{javarone_pone17}. For example, staying together may minimize predation risk, but seeking for food could also be more advantageous in groups \cite{parrish_s99,szolnoki_srep12,martinig_srep20}. Notably, such an arrangement of individuals play a decisive role in the evolution of cooperation, too, because it can reveal the sharp contrast between cooperator and defector strategies \cite{nowak_n92b}. Namely, while the defectors cannot benefit from the clustering with akin players, cooperators enjoy the close company of similar fellows and this difference is in the heart of several ways of reciprocity \cite{szolnoki_rsif15,szabo_pr07,floria_pre09,szolnoki_pre18}.

In a previous paper we studied a simple model of clustering in which female and male individuals followed a basic rule of reproduction, die and movement and demonstrated that a uniformly distributed population can transform into a state where species form a huge single cluster \cite{bazeia_epl20}. While the centre of this aggregation moves randomly its size follows a power law behavior in terms of the number of individuals in the whole population. In our present work we alter some element of former model definition slightly and demonstrate that it causes dramatic change in pattern formation. In particular, we do not apply a global restriction for the maximum value of individuals. Instead, we apply a local constraint which maximizes the number of individuals within a certain distance of parents species. Furthermore, we distinguish two characteristic lengths, which are the length of movement ($\ell_{mov}$) and the range of reproduction ($\ell_{rep}$). This extension make us possible to study the proper consequence of individual mobility on the evolutionary outcome. As result, we can observe the emergence of muti-clustered pattern in the stable phases and explore how the intensity of individual movement influences the survival of the whole population.

Beside the mentioned extensions we also study the possible consequences of collective movement, when the motion of species is not completely random, but there is a general drive in the population.  This condition can be observed in nature frequently when floating individuals are in a stream or in a river, but currents in oceans can also offer examples for similar collective drift \cite{vicsek_pr12}.  Notably, the potential impact of a drive on the final outcome is not restricted to cases when it is a given natural condition, but it could also be an artificial tool to manipulate the evolving solution of a competing system \cite{drescher_cb14, Javarone2016, Armano2017}. In this way to explore the potential link between emerging patterns and the intensity of flow could be a promising device to design a desired solution in a complex system.  

\section{The model}
\label{def}

In our minimal off-lattice model of living organisms we do not assume any particular interaction between individuals and the pattern formation is controlled solely by some very basic rules. At the beginning we release $N/2$ female and $N/2$ male individuals who are distributed randomly on a square-shaped box of linear size $L=1$. To avoid problems originated from finite size we apply periodic boundary conditions.

At each time step a single individual (active) is chosen randomly to move in a random direction with a step size $\ell_{mov}$. Furthermore, one of the two competing processes is executed with a certain probability. More precisely, an individual is chosen to die and removed from the simulation with probability $p_d$. Alternatively, a reproduction process is executed with probability $p_r=1-p_d$. In the latter case we explore the neighborhood of the focal individual within a distance $\ell_{rep}$. If there is an individual of the opposite sex within the circle then a newborn is settled whose position and gender are chosen randomly. Importantly, the new individual is settled only if the total number of individuals in the neighborhood (within a distance $R=0.1$) does not exceed a threshold value $M$. This local constraint of maximum number of individuals is a significant difference from our previous model where only a global constraint for the maximum number of total population was assumed \cite{bazeia_epl20}. Another important change comparing to our previous model is $\ell_{mov}$ and $\ell_{rep}$ could be different. In this way we can study the possible impact of individual's mobility properly.

In most of the simulations we used $p_r=0.7$ and $p_d=0.3$ but results of other probability pairs are also discussed. In the stationary state, which is typically reached within 500 generations, we have measured different quantities as described later. For the requested accuracy we have averaged our data over 500 independent runs.

Beside the random walks of individuals we have also considered biased movement when species are drifting along a specific direction collectively. Motivated by previous self-driven models \cite{vicsek_prl95,avelino_epl18, avelino_epl20} we span between the random and completely biased movement by using a driving parameter $\eta$. When $\eta=1$ individuals move randomly while at $\eta=0$ all of them move toward a specific direction. We have chosen this direction horizontally from left to right, which is connected to $\phi=0$ angle in a polar coordinate system. When the value of $\eta$ is between these two extreme cases then individuals have a limited access of choosing a direction to jump, where the $(-\phi,+\phi)$ range of available  directions is proportional to the value of $\eta$. As we already mentioned, collective movement of living species is a very common phenomenon in nature which might have a specific role on the emergence of a certain pattern. The main goal of present paper is to clarify the possible connections between individual movement and emerging global patterns.

\section{RESULTS}

\subsection{Random movement}

\begin{figure}
\centering
\includegraphics[width=13.5cm]{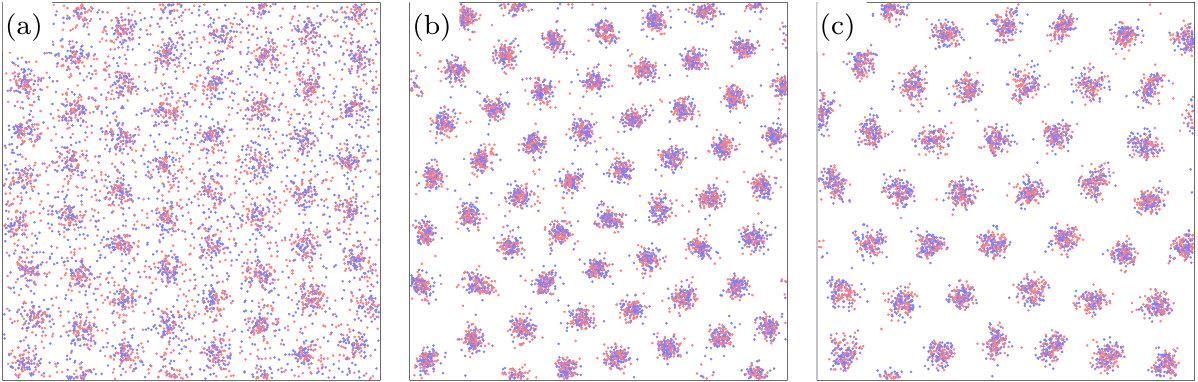}\\
\caption{Typical snapshots of the spatial distribution of females (red) and males (blues) after $500$ generations when the evolution is started from a random initial condition with $N=10^4$ individuals. Panels from (a) to (c) show the stable stationary patterns at $p_d=0.05, 0.25$ and $0.41$ respectively. For higher $p_d$ values the population goes extinct. In every cases we used the same $M=160$, $\ell_{mov}=\ell_{rep}=0.01$ parameter values for proper comparison. The applied local constraint to the maximum number of individuals result is a multi-clustered pattern where the spatial arrangement of clusters depicts a hexagonal rotation symmetry.}\label{multi-clustered}
\end{figure}

We first present some results obtained at small mobility values where individuals chose all directions with the same probability. Figure~\ref{multi-clustered} shows three typical snapshots of the population for low, intermediate and relatively high probability of dying. For proper comparison the rest of the parameters, namely the characteristic lengths of movement and reproduction and the maximum number of individuals in a neighborhood, are identical.

When the probability of die is low, shown in panel~(a), then a reasonably high number of individuals are present in the available space. As we increase $p_d$, the total number of survivors decays gradually. First those individuals become vulnerable who are far away from the centre of a cluster. But above a threshold value of $p_d$, however, all individuals go extinct and the population cannot survive no matter how large initial $N$ value was used. Evidently, the critical $p_d$ value depends on the value of $M$, which is $p_d=0.42$ for $M=160$.

The most striking difference between present results and those obtained in a former case \cite{bazeia_epl20} is the stable presence of several clusters in our present model. This is a direct consequence of the fact that we modified the restriction of maximum number of surviving individuals from a global to local one. As the previous work demonstrated, the application of a global restriction law resulted in a single-cluster final state. This behavior can be understood because the globally applied rule generated an effective interaction between competing clusters. More precisely, when parents in a specific cluster tried to give birth to a descendant then its success depended on the actual state of the population far in another cluster. Accordingly, the newborn was aborted if the global number of individual exceeded the threshold value $N$. In this way the growth of a cluster was not independent on the evolution of another competing cluster. But the lack of a newborn also implies the decay of the mentioned cluster because the alternative process, which is a random die of group members with a given probability, works freely. Evidently, an initially larger cluster has a higher chance to survive and after it can produce a huge singular aggregation to reach the maximal allowed population size. This effective interaction between the competing clusters makes the game biased because the cluster which has an initial advantage is capable to enlarge the difference and survive alone. However, when we switched the global restriction rule to a local one then clusters can evolve independently where the only restriction is to avoid too crowded spot locally. As a result, several clusters remain alive where the emerging pattern is the result of a delicate balance between the competing processes of reproduction and die.

We stress that individuals cannot survive everywhere, but just only in close neighborhood with others, which is needed to compensate natural lost. This is a fundamental condition of cluster formation where the actual value of $p_d$ determines their prevalence. The spatial distribution of cluster, however, always depicts a hexagonal rotation symmetry independently how many of them are present, as it is illustrated in Fig.~\ref{multi-clustered}. 

\begin{figure}
\centering
\includegraphics[width=13.5cm]{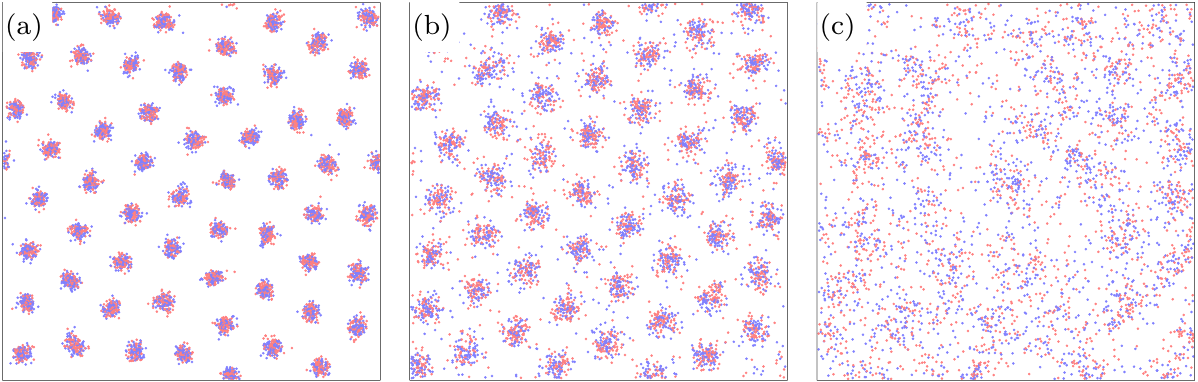}\\
\caption{Stationary snapshots for different values of individual mobility for $\ell_{mov} = 0.005$ (a), $\ell_{mov} = 0.015$ (b), and $\ell_{mov} = 0.025$ (c). The remaining parameters, $p_d=0.30$, $\ell_{rep}=0.01$, and $M = 150$, are fixed for all cases. The comparison suggests that intensive individual movement destroys the cluster ordering.}\label{l_mov}
\end{figure}

Next we explore the possible role of individual mobility on the pattern formation. In our previous model $\ell_{mov}$ and $\ell_{rep}$ had the same value, but this equality prevents us to explore the proper consequence of individual mobility. More precisely, if we enlarge both typical length scales simultaneously then there is no chance for players to leave their parents or descendants and the increase of the single characteristic length does not change the neighborhood of individuals relevantly. However, if we vary $\ell_{mov}$ unilaterally by keeping $\ell_{rep}$ fixed then individuals may arrive into a new environment where descendants or parents are not necessarily present. This extension, as it is illustrated in Fig.~\ref{l_mov}, has huge consequence on the spatial distribution of species. In the mentioned plot we show some typical patterns for the cases of minimal, mild and more intensive individual mobility. As the patterns show the contour of disjunct clusters diminishes by more intensive player's movement. 

\begin{figure}
\centering
\includegraphics[width=10.5cm]{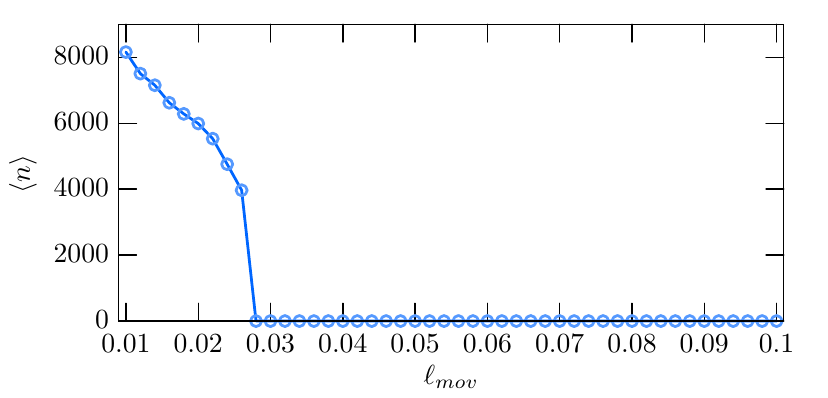}\\
\caption{Average density of the population in dependence of individual mobility for $M=160$, $\ell_{rep}=0.01$, and $p_d=0.31$. At a threshold value of $\ell_{mov} = 0.027$ the system goes extinct. It suggests that intensive individual mobility could be harmful for the whole population.}\label{destroy}
\end{figure}

Previously we argued that clustering is an essential condition for the population to survive therefore the lost of clusters can easily jeopardize the survival chance of individuals. Indeed, one may observe less individuals in panel~(c) of Fig.~\ref{l_mov} where the movement of players is the most intensive. To quantify this effect we measure the global density of individuals in dependence of $\ell_{mov}$ and the results are summarized in Fig.~\ref{destroy}. This plot confirms our previous observation, namely the average number of individuals decreases by increasing the mobility of players. Nevertheless, this decay is not continuous, but falls to zero level at a specific mobility value. This result suggests that too intensive movement of individuals could be detrimental for the survival chance of the whole population.

\begin{figure}[b!]
\centering
\includegraphics[width=10.5cm]{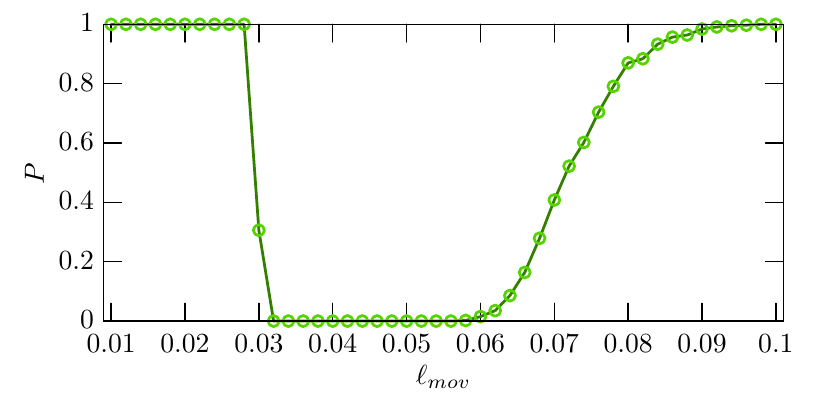}\\
\caption{Survival probability of the population ($P$) in dependence of individual mobility $\ell_{mov}$. Other parameters are $M=160$, $\ell_{rep}=0.01$, and $p_d=0.3$. Simulations were started from a random distribution of $N=10^4$ individuals and each point was calculated from $10^3$ independent runs after 500 generations. A moderate strength of individual mobility can destroy the population, but a more intensive movement is able to recover the surviving chance.}\label{recover}
\end{figure}

One may argue, however, that if the mobility is high enough then individuals may reach another cluster and in this way they get a second chance to be mated. It could be the case but only if the $p_d$ probability of die is not too high hence players survive the journey to the next aggregation. This effect is illustrated in Fig.~\ref{recover} where we present the survival probability of the whole population in dependence of the intensity of player's movement. In agreement with the previous plot shown in Fig.~\ref{destroy} slow movement of individuals does not harm the chance of survival, but longer jumps away from the original cluster members can easily kill the whole population. Increasing further the distance of jumps, hence the intensity of mobility, can give a reasonable chance to survive and it becomes practically a sure event for large $\ell_{mov}$. The significant difference between the case shown here and in our previous plot is a bit lower value of $p_d$ which allows players to arrive safe to the next harbor when movement was intensive. It is easy to understand that similar impact can also be reached by increasing the $M$ threshold value of locally maximum individuals at a fixed value of $p_d$ because in this case there are more players on the stage hence there could be someone to mate in the neighborhood which keeps the population alive.

\begin{figure}[b!]
\centering
\includegraphics[width=10.5cm]{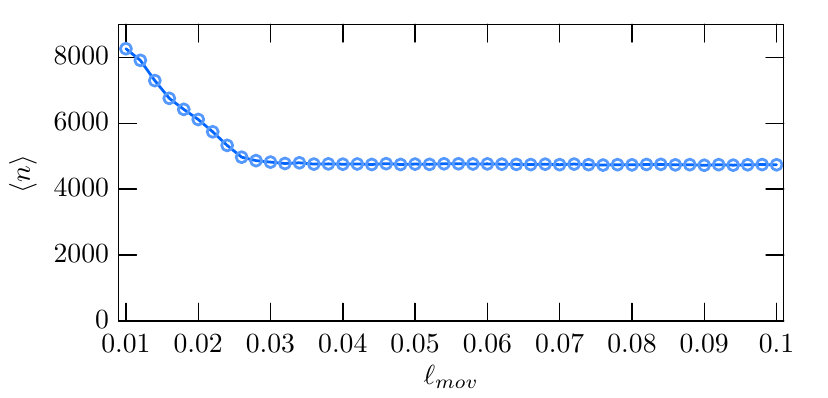}\\
\caption{Average density of the population in dependence of individual mobility for $M=160$, $\ell_{rep}=0.01$, and $p_d=0.29$. When we increase $\ell_{mov}$ from a minimal value then the concentration level decays a bit because there are players who escape from a chance of mating. But if the mobility exceeds a certain level then the possibility to find a new partner becomes viable hence the global population is not reducing further. Accordingly, the total number becomes independent of the value of $\ell_{mov}$.}\label{saturate}
\end{figure}

If we accept the above described explanation how pattern is forming in the presence of player's mobility then it is a straightforward consequence that by decreasing $p_d$ further we can avoid the extinction of the population even for more intensive individual movement. Indeed, the results plotted in Fig.~\ref{saturate} confirm our claim where the population does not go extinct no matter how intensive mobility was applied. Naturally, by increasing $\ell_{mov}$ from a minimal value does lower the average concentration because some player escapes from the chance of mating, but eventually they will find a new partner to produce new members who can compensate the natural loss of the population. In this way the strength of mobility becomes second-order important and the average concentration level saturates which value depends exclusively on the values of $M$ and $p_d$.

\subsection{Collective driven movement}

In the rest of this paper we consider the case when the players movement are not completely random, but follow a kind of ordering where  limited jumping directions are allowed only. Here the value of $0 \le \eta \le 1$ determines the available portion of jumping directions from the whole $[-\pi,+\pi]$ interval.

\begin{figure}[b!]
\centering
\includegraphics[width=10.5cm]{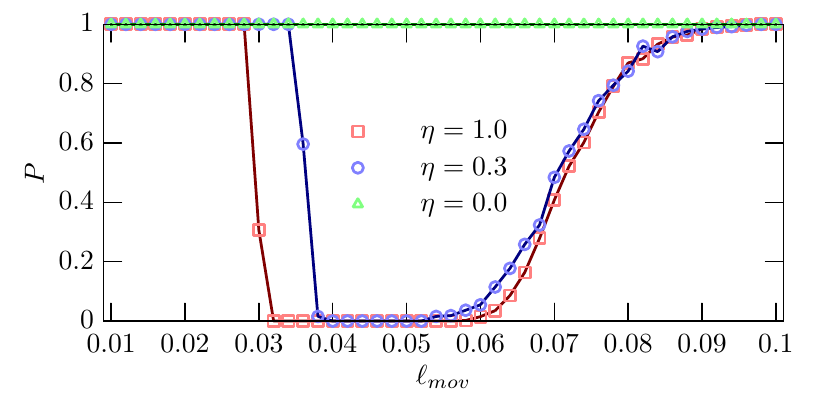}\\
\caption{Survival probability of the population ($P$) in dependence of individual mobility $\ell_{mov}$ for different values of drive. Here $\eta$ parameter determines the portion of available jumping directions for individuals, where $\eta=1$ denotes random jumps to the whole $[-\pi,+\pi]$ range, while $\eta=0$ marks a completely reduced, hence ordered, movement along the $x$-axis exclusively. Other parameters are $M=160$, $\ell_{rep}=0.01$, and $p_d=0.3$. This comparison suggests that by increasing the ordering of individual movement we can improve the survival chance of the population even in the case where intensive mobility would dictate extinction in the random movement case.}\label{drive_P}
\end{figure}

Our first principal observation is that restricted movement alone is capable to improve the chance of survival if all other circumstances remain unchanged. This effect is summarized in Fig.~\ref{drive_P} where we plotted the survival probability as a function of individual mobility for different values of drive. Importantly, all remaining parameters are equal for all cases. As we already noted, certain strength of mobility can be harmful for the population if players jump randomly in all directions. But as we reduce the range of available jumping directions the survival chance becomes better for more and more mobility values. In the extreme case, when the individual movement becomes completely biased then the population survives irrespectively of the value of $\ell_{mov}$.

\begin{figure}
\centering
\includegraphics[width=13.5cm]{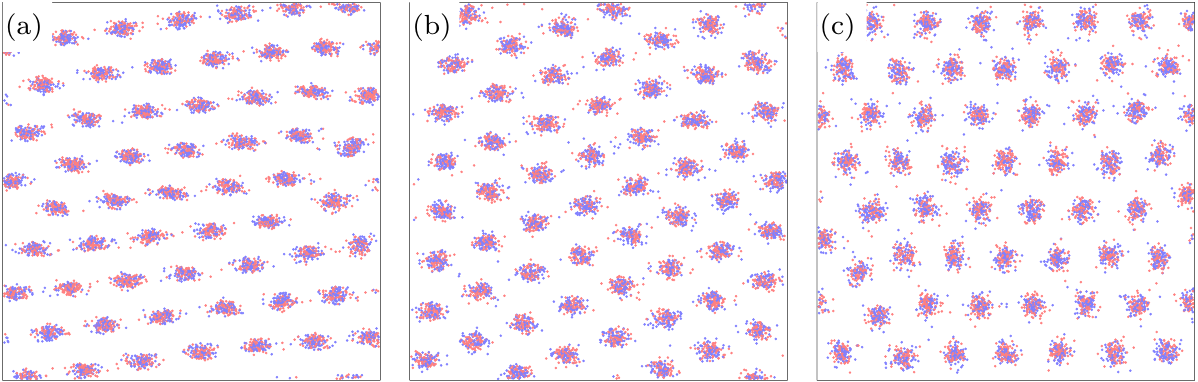}\\
\caption{Typical spatial distribution of individuals for different driving values of $\eta$. Panels from (a) to (c) show snapshots for $\eta=0, 0.2$, and $0.9$ 
respectively. For proper comparison all other parameter values, namely $M=160$, $p_d=0.30$, and $\ell_{mov}=\ell_{rep}=0.01$, are fixed for all cases. As we increase the drive, means reduce the range of jumping direction, the shape of the clusters becomes squeezed and the original hexagonal symmetry of spatial arrangement eventually becomes a simple stripe-like pattern where the stripes are along the direction of the drive.}\label{squeezed}
\end{figure}

In the following we present some representative snapshots of spatial arrangement of individuals for different drive values. The key point is summarized in Fig.~\ref{squeezed} where we plotted both the extreme cases and a sample for an intermediate drive value. In agreement with our previous observations the individuals are gathered in clusters whose arrangement show the mentioned hexagonal symmetry when players move randomly. As we introduce a drive and increase its value then the shape of the cluster becomes squeezed. Furthermore, the relative spatial positions of these clusters also change and the new order breaks the original hexagonal symmetry. Instead, squeezed clusters order into stripes which are parallel to the direction of the drive.

The above described pattern formation helps to understand why a collective driving promotes the survival of the whole population. When the individual movement is not random, but tends toward a specific direction then it keeps traveling individuals together, which always offer a higher chance of mating. Therefore in this case the negative consequence of individual mobility becomes not as crucial as in the previously discussed section. Evidently, the stronger the drive the smaller the spreading of former partners, hence movement plays no fundamental role anymore.

As a consequence of the above described phenomenon, in the $\eta=0$ limit we cannot talk about clusters anymore, but instead parallel stripes. It is a natural expectation that the orientation of the stripes is parallel to the direction of the drive. In other words, the flow of individuals generates a characteristic spatial patterns. We note that a similar selection from possible patterns by an externally flow was already observed in driven diffusive systems earlier \cite{szabo_pra90,szabo_prb92}. But in the latter case the short range interaction among particles established the potential ordered state of the system. In our present case, however, there is no similar kind of interaction among players, they interact via the dynamical rules of die and reproduction only.

\begin{figure}
\centering
\includegraphics[width=13.5cm]{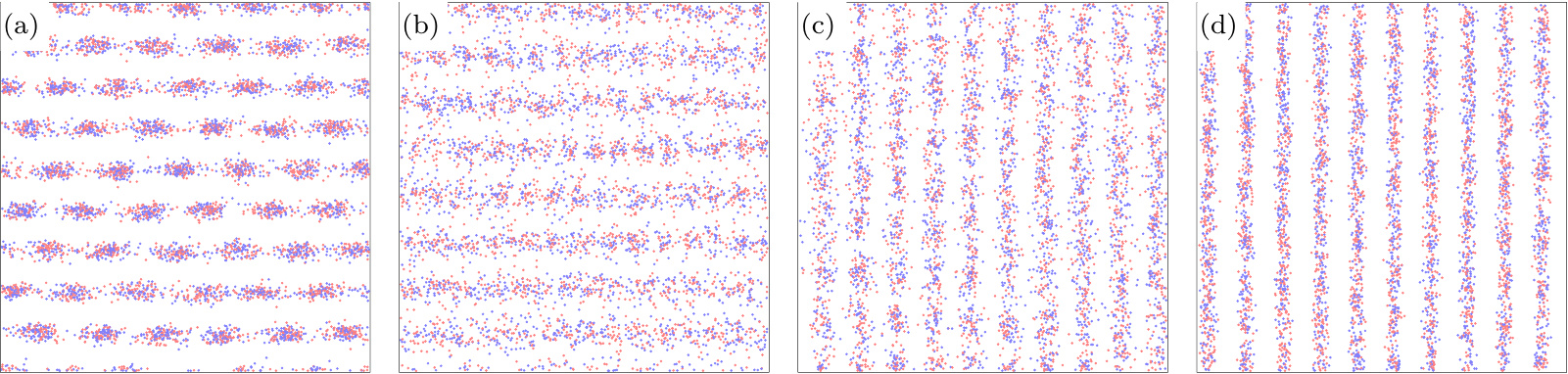}\\
\caption{Typical spatial distribution of individuals for different strength of movement for unidirectional drive at $\eta = 0.1$ %when all players jump from left to right. 
Panels from (a) to (d) show snapshots for $\ell_{mov}=0.020$, $0.070$, $0.095$, and $0.100$ respectively. Other parameters, such as $M=160$, $p_d=0.30$, and $\ell_{rep}=0.01$, are fixed for all cases. When we increase the intensity of players movement the orientation of stripes flips perpendicular to the direction of drive.}\label{flip}
\end{figure}

Quite surprisingly, the orientation of stripes does depend on the intensity of individual movement. More precisely, until a threshold level of $\ell_{mov}$ the stripes are parallel to the flow of players, but above this specific value the direction of the stripes flips and they become perpendicular to the original ordering. This phenomenon is illustrated in Fig.~\ref{flip}, where we present some typical snapshots for different values of individual jumps. Here the first two panels were taken at small and moderate $\ell_{mov}$ values, while the remaining two panels show the pattern for more intensive individual movement. Importantly, all other parameter values, like $\eta$, $M$, $\ell_{rep}$, or $p_d$, are equal for all cases. One may notice that the contour of stripes becomes sharper as we leave the threshold value, which happens for low and large $\ell_{mov}$ values, while the stripes become blurred in the vicinity of the transition point. The number of stripes may also change because, as we pointed out earlier, the total number of individual becomes independent of the strength of movement and it practically depends on the value of $p_d$. Hence if we have wider stripes then their numbers should decrease simultaneously.

\begin{figure}
\centering
\includegraphics[width=13.5cm]{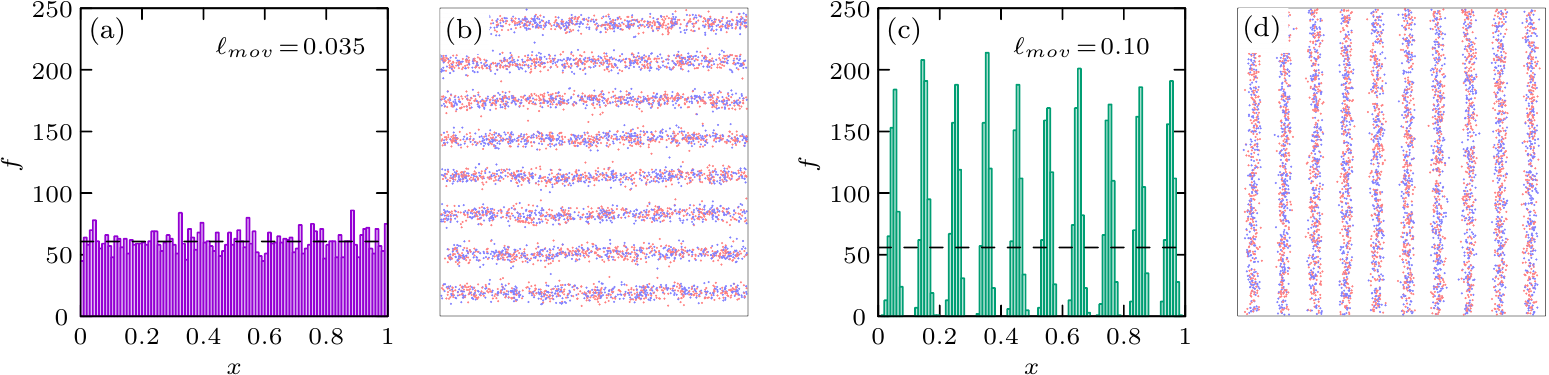}\\
\caption{Typical spatial distribution of individuals from both sides of the transition point. Panel~(b) shows the pattern at $\ell_{mov}=0.035$, while panel~(d) at $\ell_{mov}=0.100$. To distinguish them we measured the frequency of individuals in dependence of their horizontal positions and the related histograms are shown in panel~(a) and (c) respectively. As the histograms show, when stripes are horizontal then the profile function $f$ fluctuates slightly around an average level, which is marked by dashed line. In case of vertical stripes, however, there are huge differences in the profile function as we vary the horizontal coordinate. This makes possible to introduce an order parameter which measures the variance of the profile function.}\label{order_parameter}
\end{figure}

To quantify the ordering process we utilize that fact that when stripes are parallel to the drive then the horizontal position of individuals has no relevance. On the contrary, when stripes are perpendicular to the drive then there is a certain ordering in the horizontal positions of players: there are crowded and deserted intervals as we vary horizontal coordinates. This difference makes us possible to introduce an order parameter which measures the difference between the horizontal and vertical ordering. In particular, we measure the horizontal positions of all individuals and make a histogram of the frequency of their values. Evidently, the number of disjunct intervals should exceed the number of stripes, but if we choose the former large enough then the profile function of the histogram will not change relevantly. As we argued above, there is a huge difference between the profile functions if the ordering of the stripes is horizontal or vertical. The difference is summarized in Fig.~\ref{order_parameter} where we present two representative patterns and the related histograms. As panel~(a) illustrates, the horizontal positions of players show a uniform distribution when stripes are horizontal and the related frequencies fluctuates slightly around an average value. When the stripes are vertical, however, the mentioned frequency profile becomes strongly periodic where large and zero values occur periodically, which reflects a horizontal ordering faithfully. Therefore, if we measure the variance of the frequency profiles then we get a small or high $\sigma_f$ value depending on the horizontal or vertical directions of the stripes. 

\begin{figure}
\centering
\includegraphics[width=10.5cm]{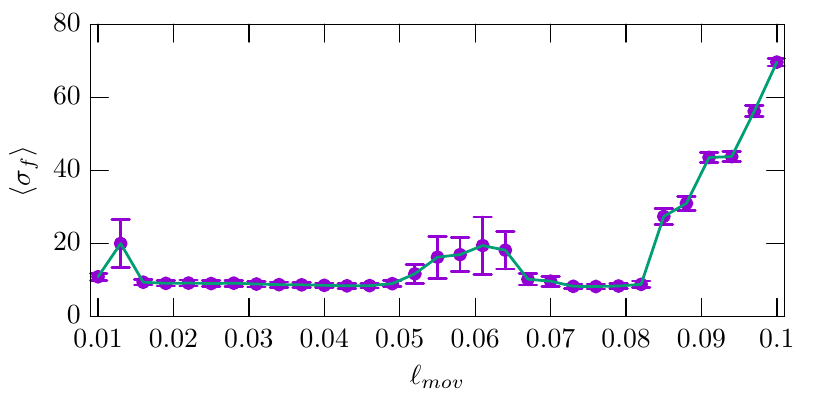}\\
\caption{The order parameter which characterizes the flipping transition in dependence of $\ell_{mov}$ at $\eta=0.1$, $M=160$, $p_d=0.30$, $\ell_{rep}=0.01$. Note that $\sigma_f$ is not completely zero even for horizontal ordering because there is some noise driven fluctuation in the $f$ profile as it is shown in panel~(a) of Fig.~\ref{order_parameter}. Furthermore, a smaller bulge can also be detected in the $0.052<\ell_{mov}<0.066$ interval where the stripes are still horizontal but there are some empty gaps in them which makes the mentioned profile fluctuating. For $\ell_{mov}<0.08$, however, the tendency is clear that is a mark of the transition in the orientation of the stripes.}\label{flip_op}
\end{figure}

Indeed, if we measure $\sigma_f$ in dependence of $\ell_{mov}$ then we can detect quantitatively the flipping transition we reported above. This function is shown in Fig.~\ref{flip_op} where we reduced all $\sigma_f$ values by the value obtained at small $\ell_{mov}$ where horizontal stripes are present. In this way the positive value and the magnitude of the reduced $\sigma_f$ characterizes the measure of ordering into vertical stripes. When this value is low then the difference of ordering  between the two directions is not relevant, but when the mentioned value is high then we have a decent vertical ordering of stripes. This plot identifies clearly a transition point when the flip of ordering starts. Evidently the critical point of this transition depends on the value of $M$ and $p_d$, but it is robust and can always be detected when we apply a biased movement of individuals.

\section{DISCUSSION AND CONCLUSIONS}

In this work we have explored the possible consequences of individual mobility on the pattern formation in a simple model of living organisms. Our principal goal was to clarify whether the intensity of movement is crucial and check its impact on the collective behavior in various conditions. We should stress that our model is very simple and general, hence it could be adopted for wide range of realistic systems \cite{durrenberger_jsb91}.

As a key observation, we pointed out that a multi-cluster pattern emerges if we leave the global constraint of maximum number of living individuals and replace it by a locally applied restriction. Despite of the fact that we used a numerically demanding off-lattice simulation, the resulting pattern can be described by a hexagonal symmetry in a broad range of parameter values. This behavior justifies the widely used assumption which states that triangular host lattice is an adequate topology to simplify biological system during a discrete-modeling \cite{mouritsen_84,newman_99,szolnoki_pre04}.

We also found that intensive random individual mobility could be harmful for the whole population because it reduces the occasions when partners with opposite sex meet and produce a descendant. However, if the players movement is intensive enough then the mentioned negative consequence can be fixed and the stability of the population is restored.

Furthermore, when the individual motion is systematic and a collective macroscopic movement emerges due to an external bias then the above mentioned negative consequence of intensive mobility cannot be observed. Hence we can conclude that a flow of participants stabilize the living conditions of an organism where departing from others would easily jeopardize the whole community otherwise.

The flow towards a specific direction has an additional role on the emerging pattern, too, because it alters the spatial symmetry of the cluster distribution. In particular, the original hexagonal symmetry is replaced by a two-fold symmetry where individuals are organized into parallel stripes. Interestingly, the direction of the stripes can be flipped if the individual movement exceeds a threshold value. Above it the perpendicular ordering becomes typical where the transition can be characterized by an appropriately chosen order parameter.

We would like to stress that the above mentioned external bias may not necessarily be a given natural condition, but could also be a human controlled arrangement. The latter makes possible to utilize the flow of individual to design and manipulate emerging patterns. Such efforts have already been applied in engineered structure to study microfabricated habitats \cite{keymer_pnas08}, but fluid advection was also used to study emerging patterns in interacting microbial communities \cite{uppal_elife18}. We hope that our present work will be useful and inspiring for future experimental works.

\vspace{0.5cm}

%\section*{ACKNOWLEDGMENTS}

B.F.O. and M.V.M. thank CAPES - Finance Code 001, Funda\c c\~ao Arau-c\'aria, and INCT-FCx (CNPq/FAPESP) for financial and computational support. D.B. acknowledges Conselho Nacional de Desenvolvimento Cient\'\i fico e Tecnol\'ogico (CNPq, Grants nos.  303469/2019-6 and 404913/2018-0) and Para\'\i ba State Research Foundation (FAPESQ-PB, Grant no. 0015/2019) for financial support.

\bibliographystyle{elsarticle-num-names}

%\bibliography{ref}

\end{document}